# Using State Infection Conditions to Detect Equivalent Mutants and Speed up Mutation Analysis

– Presented at Dagstuhl Seminar 13021: Symbolic Methods in Testing –

René Just, Michael D. Ernst
Computer Science and Engineering
University of Washington
Seattle, WA, USA
{rjust, mernst}@cs.washington.edu

Gordon Fraser
Department of Computer Science
University of Sheffield
Sheffield, UK
gordon.fraser@sheffield.ac.uk

*Abstract*—Mutation analysis evaluates test suites and testing techniques by measuring how well they detect seeded defects (mutants). Even though well established in research, mutation analysis is rarely used in practice due to scalability problems — there are multiple mutations per code statement leading to a large number of mutants, and hence executions of the test suite. In addition, the use of mutation to improve test suites is futile for mutants that are equivalent, which means that there exists no test case that distinguishes them from the original program.

This paper introduces two optimizations based on state infection conditions, i.e., conditions that determine for a test execution whether the same execution on a mutant would lead to a different state. First, redundant test execution can be avoided by monitoring state infection conditions, leading to an overall performance improvement. Second, state infection conditions can aid in identifying equivalent mutants, thus guiding efforts to improve test suites.

## I. INTRODUCTION

Mutation analysis is a fault-based technique that assesses the quality of a test suite by measuring the number of seeded defects (*mutants*) that the test suite reveals (*kills*). However, the number of such mutants is typically huge, making the technique unscalable for large programs. Furthermore, some of the mutants are equivalent, meaning that there exists no test case that can distinguish the mutant from the original version, such that attempts to generate tests to kill them are futile. Unscalability and the equivalent mutant problem limit the applicability of mutation analysis in practice.

This paper addresses these problems by exploiting state infection conditions. A state infection condition determines whether a test execution would lead to a different state on a mutated version of the program. If this condition is true, then the mutant differs from the original version at some point during its execution. Although these conditions are well known, there is further potential to use them to address the problems of scalability and equivalent mutants.

To avoid unnecessary test executions and improve scalability, state-of-the-art mutation analysis tools gather coverage information in a preprocessing step. A test execution is skipped if it would not reach the mutated code location. This can be further improved by checking state infection conditions during the preprocessing step. If a test does not achieve state infection on a mutant, then the test cannot kill it, and hence there is no need to execute the test for this mutant. Consequently, state infection conditions reduce the overall number of test executions. Note that a mutant that does not exhibit state infection either indicates a weakness of the test suite or it is equivalent with regard to all possible tests.

One use of mutation analysis is to improve a test suite by adding new tests that kill each unkilled mutant. Yet, a mutant might be unkilled because it is equivalent, i.e., no test exists that would kill it. In this case any attempt to generate a test to kill it is futile. This paper uses state infection conditions to query constraint solvers. If such a state infection condition is unsatisfiable, then this reveals an equivalent mutant.

In detail, the contributions of this paper are:

- Section II empirically compares the efficiency improvements by exploiting two necessary conditions to kill a mutant, namely mutation coverage and state infection.
- Section III presents a new approach based on state infection conditions and symbolic execution that uses constraint solvers to identify equivalent mutants.
- Section IV provides a case study of applying this approach to identify equivalent mutants in an example application.

## II. STATE INFECTION VS MUTATION COVERAGE

Mutation analysis executes a test suite on a set of mutants to measure the test suite's ability to kill the mutants. For some mutants, it can be known in advance that a test suite cannot kill them. In particular, if a test suite does not even execute the mutated code, then no test needs to be executed on the corresponding mutant. This is a simple yet powerful optimization, as the mutation coverage of the test suites of real-world applications is often small, sometimes even below 50% [6].

Even if a mutant is covered, it cannot be killed unless its execution additionally causes a state infection [1]. For expression-based mutations, such as the replacement of operators or logical connectors, state infection means that the resulting value of the mutated expression differs from the value of the original expression. For example, the expression `a>b`

TABLE I
COMPARISON OF MUTATION COVERAGE WITH STATE INFECTION CONCERNING THE NUMBER OF EXECUTED MUTANTS AND THE TOTAL RUNTIME OF THE MUTATION ANALYSIS PROCESS.

|  | Total number of mutants | | Mutation coverage | | State infection | | | |
|---|---|---|---|---|---|---|---|---|
|  | Generated | Killed | Executed mutants | Total runtime* | Executed mutants | | Total runtime* | |
| itext | 126,781 | 4,876 | 18,170 | 315 | 15,718 | ( -13%) | 279 | ( -11%) |
| trove | 72,959 | 3,523 | 6,137 | 49.6 | 5,242 | ( -15%) | 36.2 | ( -27%) |
| jfreechart | 68,503 | 13,132 | 36,298 | 582 | 31,165 | ( -14%) | 470 | ( -19%) |
| joda-time | 23,781 | 14,629 | 19,602 | 214 | 17,913 | ( -8.6%) | 136 | ( -36%) |
| commons-lang | 21,056 | 14,780 | 20,196 | 48.9 | 18,120 | ( -10%) | 39.1 | ( -20%) |
| jdom | 10,800 | 8,187 | 10,266 | 108 | 9,780 | ( -4.7%) | 82.1 | ( -24%) |
| commons-io | 7,319 | 3,288 | 4,255 | 4.20 | 3,686 | ( -13%) | 2.80 | ( -33%) |
| jaxen | 7,132 | 2,078 | 4,679 | 386 | 3,608 | ( -23%) | 221 | ( -43%) |
| numerics4j | 5,437 | 3,500 | 5,209 | 1.49 | 4,972 | ( -4.5%) | 1.21 | ( -19%) |
| avg |  |  |  |  |  | ( -12%) |  | ( -26%) |

*Total runtime of mutation analysis in minutes (includes preprocessing).

and the mutated version `a>=b` only evaluate to a different outcome if `a` equals `b`. This is exploited in *weak mutation* testing, where a mutant is considered to be killed if it achieves state infection. In contrast, *strong mutation* additionally requires a propagation of the state infection to an observable output or assertion failure. A key advantage of state infection is that it can be determined without executing any mutants, but by only checking state infection conditions during the execution of the original version. Therefore, we propose to use state infection also for strong mutation testing to determine whether a test needs to be executed for a certain mutant.

In comparison with mutation coverage, the state infection analysis causes an additional overhead when executing the original version to determine whether this execution fulfills the infection conditions. To compare the costs and benefits of determining state infection, we conducted an empirical study using the MAJOR [7] mutation analysis tool. Table I gives the results for the nine open-source projects, which we investigated by executing their corresponding JUnit test suites. When compared to mutation coverage, the state infection approach decreases the number of mutants that need to be analyzed by 12% and the total runtime of the mutation analysis process by 26%, on average. Note that the total runtime includes the preprocessing time, which is several orders of magnitude smaller than the total runtime, and hence negligible.

## III. CONSTRAINTS ON EQUIVALENT MUTANTS

Identifying equivalent mutants is crucial, not only for efficiency reasons but also to avoid futile effort to generate tests to kill them. As discussed in Section II, exploiting the state infection as a necessary condition to kill a mutant improves the runtime of the mutation analysis process. However, the question remains whether the excluded mutants that fulfill the reachability but not the state infection condition are either equivalent or identify a weakness in the test suite.

This section presents a new and sound approach to weed out equivalent mutants focusing on those that are covered but do not exhibit state infection. Figure 1 visualizes the complete workflow with the following steps: First, it determines, in the preprocessing step, the sets of covered and state-infected mutants. Note that this step only involves a single execution of the original program, which is instrumented to provide all necessary information. The preprocessing step also computes all covered mutants that do not manifest a state infection. The corresponding set of these potentially equivalent mutants,

```
1  public int classify(int a, int b, int c){
2    if( a<=0 || b<= 0 || c<=0 ){
3      return INVALID;
4    }
5
6    if( a==b && a<=c ){
7      return EQUILATERAL;
8    }
9    ...
10 }
```

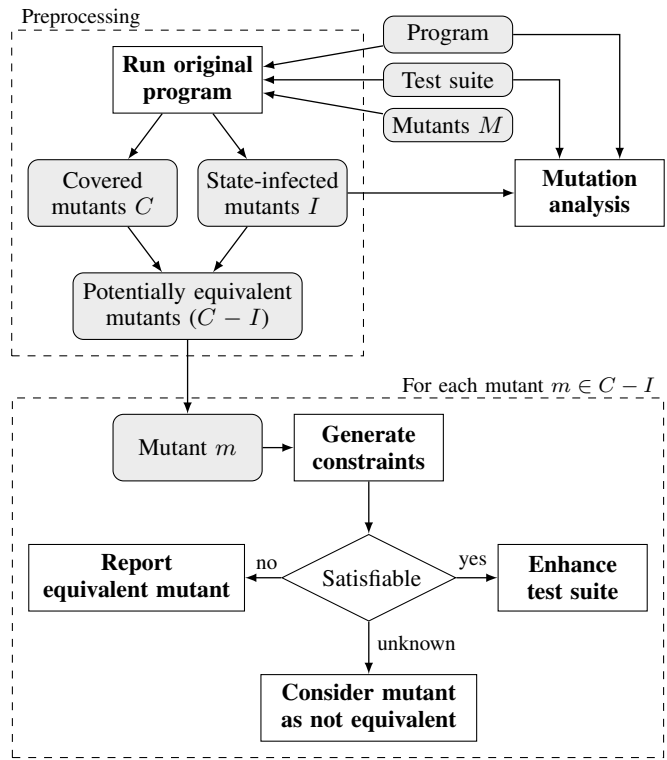

Fig. 1. Combining dynamic analysis (Preprocessing) and constraint solving to detect equivalent mutants. Note that $I \subseteq C \subseteq M$.

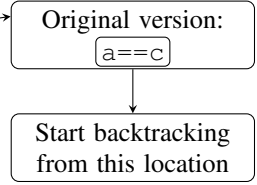

Original version: a==c

Start backtracking from this location

⑥  (a==c) != (a<=c)        } infection
     AND
⑥  (a==b)
     AND                   } reachability
④  !(a<=0 || b<=0 || c<=0)

Fig. 2. Determination of state infection conditions under which the expression value of the mutant differs from the original version. A circle indicates the line number for which the corresponding condition must hold.

represented by $C - I$, is obtained by subtracting the set of state-infected mutants from the set of covered mutants.

For each of these mutants, the workflow thereafter generates the constraints that have to be fulfilled to cause the state infection by backtracking the path from the mutated location to the entry point of the corresponding method. Intuitively, the necessary constraints are composed of the reachability conditions (i.e., backward symbolic execution) and the requirement that the mutant's value differs from the original one.

Finally, it employs a constraint solver to decide whether the extracted constraints, which would yield a state infection, can be satisfied. If the constraint solver reveals that the extracted constraints are unsatisfiable, the mutant is equivalent for the execution path chosen by the test, and does not need to be considered again for a test case that follows the same control flow. If the state infection has been shown to be unsatisfiable for all paths reaching the mutant, then this proves that the mutant is equivalent and can be completely removed from the set of mutants. On the other hand, if the solver provides a solution to the given constraints, the corresponding values can be used to enhance the test suite. Note that the state infection conditions may not be determinable (e.g., for loops) or the constraint solver may not be able to solve them (e.g., unsupported arithmetic). In this case, the decision is *unknown* and the corresponding mutant is treated as a killable mutant.

Figure 2 visualizes the determination of the state infection condition for an example code fragment and also gives the corresponding conditions that are necessary to reach the mutant `a<=c` and to cause it to evaluate to a different value than the original version `a==c`.

## IV. CASE STUDY: EQUIVALENT TRIANGLE MUTANTS

We applied the approach of Section III to the well-known triangle example, an application that classifies triangles. In order to obtain an unbiased initial test suite (*T1*) for the triangle program, we employed EVOSUITE [2] to generate a test suite that satisfies 100% branch coverage. This generated test suite consists of 14 test cases and also achieves 100% mutation coverage, meaning that all mutants are reached and executed. However, only 108 of the 125 executed mutants yield a state infection. As previously stated, the strong mutation analysis process only executes these 108 mutants since the remaining 17 mutants cannot be killed. Eventually, the generated test suite kills 92 out of 108 state-infected mutants.

TABLE II
NUMBER OF EQUIVALENT AND KILLED MUTANTS OUT OF 17 INVESTIGATED AND POTENTIALLY EQUIVALENT MUTANTS.

| Number of mutants | Solver decision | Approach decision | Killed by new test case | Precise result |
|---|---|---|---|---|
| 8 | unsat | equivalent | no | equivalent |
| 8 | sat | killable | yes | killable |
| 1 | sat | killable | no | equivalent |

### A. Equivalent Mutants due to Unsatisfiable Constraints

Since our approach aims at detecting equivalent mutants, which are covered but cannot be killed due to unsatisfiable state infection conditions, we focus on the 17 mutants that potentially belong to this category of equivalent mutants.

For each of the 17 mutants, we manually performed backward symbolic execution on all paths to the method entry and included the condition that forces the mutated expression to take a value that differs from the original one. The Z3 constraint solver then solved the constraints for each mutant. Z3 determined that 8 out of 17 constraints are indeed unsatisfiable. Hence, the corresponding mutants are equivalent and could be removed. Moreover, we used the computed solutions for the constraints of the remaining 9 mutants to form additional test cases, resulting in the test suite *T2*. These new tests not only infected the state as required by the constraints, but also killed further 8 of the 9 remaining mutants, thus leaving only one undetected. Table II summarizes all results of this analysis.

### B. Enhanced State Infection for Composed Expressions

So far, we considered the satisfiability of the state infection conditions on the mutated expression, and these conditions turned out to be quite powerful in detecting equivalent mutants. However, executing the test suite *T2* still results in 14 unkilled yet state-infected mutants. These mutants can either indicate a weakness in the test suite, or they are equivalent despite locally infecting the state. One possible case where this might happen is if the mutation is a subexpression of a composed expression. To investigate whether this is the case, we strengthen the state infection conditions for such mutants. Consider an example derived from the triangle application with the inputs `a=1`, `b=2`, `c=3`, and the following composed expressions:

$$\boxed{(a+b)} > c \quad \text{(original version)}$$

$$\boxed{(a*b)} > c \quad \text{(mutated version)}$$

TABLE III
COMPARISON OF THE INITIAL TEST SUITE (T1) AND THE TEST SUITES (T2 AND T3), WHICH ENHANCE T1 BASED ON STATE INFECTION FOR THE MUTATED EXPRESSION (T2) AND THE OUTERMOST EXPRESSION (T3).

| Test suite | | Number of mutants out of 125 | | |
|---|---|---|---|---|
| Set | Number of tests | Covered | State-infected | Killed |
| T1 | 14 | 125 | 108 | 92 |
| T2 | 22 | 125 | 117 | 103 |
| T3 | 33 | 125 | 117 | 116 |

For the given inputs, the mutant indeed manifests a state infection, (i.e., `(a*b) != (a+b)`). However, this local state infection has no effect on the outer expression since both versions evaluate to `false` (i.e., `(a*b)>c == (a+b)>c`). Hence, the mutant cannot be detected with the corresponding inputs, even though it induces a locally infected state.

Therefore, we also determined the constraints to infect the outermost expression for all remaining live mutants, in order to assess the improvements of applying this enhanced state infection. By analyzing the remaining 14 mutants, not killed by *T2*, we found that all of the constraints for the outermost expressions are satisfiable, i.e., no additional equivalent mutant was detected. This suggests that focusing on state infection directly on the mutated expression is sufficient to detect the majority of the equivalent mutants. However, the case study program and the corresponding number of mutants is small, so we need to investigate this matter for larger programs.

Furthermore, the additional tests derived from the computed solutions for the state infection of the outermost expression, resulting in the test suite *T3*, were able to kill 13 of the 14 mutants. The remaining live mutant is again the one already reported in Section IV-A. By manually inspecting this live mutant, we discovered that it is equivalent due to unsatisfiable constraints beyond the state infection of its outermost expression. Table III summarizes the results for the initial test suite *T1*, and the enhanced test suites *T2* and *T3* ($T1 \subset T2 \subset T3$). Note that some derived tests detect more than one mutant, explaining why the increase of the number of tests is smaller than the increase of the number of killed mutants. Concerning the number of killed mutants, using state infection for the outermost expression seems to be adequate for generating tests that not only infect the state but also kill the mutant.

## V. RELATED WORK

The idea of state infection conditions goes back to DeMillo et al. [1], who identified reachability, infection, and propagation as the three necessary conditions to kill a mutant. Their state infection conditions have since been used to drive test generation (e.g., [1], [4]). In this context, the conditions have also been used to avoid redundant test executions during fitness evaluations in a search-based test generation approach [3]. In contrast, our approach uses state infection during mutation analysis, and avoids executing tests on mutants if they do not infect the state. This extends the common strategy to measure mutation coverage to reduce the number of test executions.

A commonly assumed workflow in mutation analysis is that the developer applies the analysis to an existing test suite, and uses the results to further improve the test suite. If a mutant is not even covered, then this indicates that a new test case is required, while a mutant that is covered but not killed might identify a weak test oracle (e.g., a missing test assertion). Schuler et al. [9] proposed to measure the *impact* of a mutant in terms of invariant violations or changes in code coverage. The impact of a mutant is correlated with its probability of being equivalent, such that improvement efforts should start with the mutants with highest impact. However, covered mutants without state infection have no impact per definition, leaving the question of what to do about these. They might be equivalent, or they might just be very difficult to kill, such that a killing test would greatly improve the test suite. Our approach addresses precisely this neglected category of mutants.

Besides impact analysis [9], there have been a number of proposed techniques to address the equivalent mutant problem [5]. In particular, Nica and Wotawa [8] convert programs to a constraint representation, and determine equivalence by solving a constraint system consisting of original program and mutant. In contrast, our approach only represents individual program paths, and uses state infection conditions to check whether mutants can infect the state, which results in simpler constraints. State infection is a necessary but not sufficient condition for non-equivalence, e.g., the state difference may not propagate to an observable output. In our study, only one out of 9 triangle mutants was equivalent despite state infection, suggesting that this is an acceptable trade-off in order to simplify the resulting constraint systems.

## VI. CONCLUSIONS

Despite significant progress, mutation analysis still suffers from problems concerning scalability and equivalent mutants. This paper has presented two new optimizations to the mutation analysis process. First, computing state infection, a necessary condition to kill a mutant, during the initial test execution can significantly reduce the overall number of mutants that need to be analyzed. Second, as a further application of the state infection conditions, a constraint solver can detect locally equivalent mutants. It identified 8 out of 9 equivalent mutants in our case study. Furthermore, by augmenting the test suite with the computed solutions of the constraint solver, we were able to kill 24 previously undetected, yet covered mutants.

This paper has presented the ideas and some initial results, but the bulk of implementation remains to be done in order to perform large scale experiments.